\documentclass[aps,pre,twocolumn,a4paper,10pt,notitlepage,footinbib,superscriptaddress,longbibliography]{revtex4-1}
\usepackage[english]{babel}
\usepackage{lmodern}
\usepackage[utf8]{inputenc}
\usepackage{endnotes}
\usepackage{amssymb,amsmath,amsfonts}
\usepackage{textcomp}
\usepackage{url}

\usepackage{graphicx}% Include figure files
\usepackage{dcolumn}% Align table columns on decimal point
\usepackage{bm}% bold math
\usepackage{xcolor}

\begin{document}

\title{Quantum wave representation of dissipative fluids}

\author{L. Salasnich}
\affiliation{Dipartimento di Fisica e Astronomia ``Galileo Galilei'' and Padua QTech Center, Università di Padova - Via Marzolo, 8, Padova, 35131, Italy}
\affiliation{Sezione di Padova, INFN, - Via Marzolo, 8, Padova, 35131, Italy}
\affiliation{Istituto Nazionale di Ottica, CNR - Via Carrara, 1, Sesto Fiorentino, 50019, Italy}
\author{Sauro Succi}

\affiliation{Center for Life Nano Science@La Sapienza, Istituto Italiano di Tecnologia, 00161 Roma, Italy}
\affiliation{Institute for Applied Computational Science, John A. Paulson School of Engineering and Applied Sciences, Harvard University, Cambridge, Massachusetts 02138, USA}
\affiliation{Istituto per le Applicazioni del Calcolo, CNR, Via dei Taurini 19, Rome, 00185, Italy}
\author{Adriano Tiribocchi}
\affiliation{Istituto per le Applicazioni del Calcolo, CNR, Via dei Taurini 19, Rome, 00185, Italy}
\affiliation{INFN "Tor Vergata" Via della Ricerca Scientifica 1, 00133 Roma, Italy}

\begin{abstract}
We present a mapping between a Schr\"odinger equation with a shifted non-linear potential and the Navier-Stokes equation. Following a generalization of the Madelung transformations, we show that the inclusion of the Bohm quantum potential plus the laplacian of the phase field in the non-linear term leads to continuity and momentum equations for a dissipative incompressible Navier-Stokes fluid. An alternative solution, built using a complex quantum diffusion, is also discussed. The present models may capture dissipative effects in quantum fluids, such as Bose-Einstein condensates, as well as facilitate the formulation of quantum algorithms for classical dissipative fluids.     
  \end{abstract}

\maketitle  

\section{Introduction}\label{sec1}

Formal analogies between quantum and fluid mechanics have been noticed since the early days of quantum physics.
Particularly, Madelung first showed that by expressing the complex wavefunction in eikonal form $\Psi = R e^{iS/\hbar}$ (where $R$ is the amplitude and $S$ is the action), the real and imaginary part of the Schr\"odinger equation turn
into the equations (continuity plus momentum) of a perfect (non-dissipative) compressible fluid whose quantum nature  is revealed by the presence of the so-called quantum potential $Q$, which has no classical counterpart \cite{Madelung1,Madelung2}. 
The Madelung approach is useful in many respects: from the theoretical standpoint  it is conducive to de Broglie's pilot wave formulation \cite{debr1,debr2} and lately  to Bohm's theory of hidden variables \cite{bohm1,bohm2,messiah}.
Although both formulations were largely overshadowed by the Copenhagen interpretation, modern developments in quantum physics, particularly the experimental demonstration of
non-locality as first postulated by John Bell \cite{bell}, are somewhat vindicating their merits.

Besides fundamentals, the quantum fluid formalism is often useful for the interpretation of hydrodynamic quantum analogs, which explore the capability of classical systems (such as walking droplets \cite{couder,bush}) to display behaviors akin to those arising in quantum mechanics, for the study of cosmic fluids \cite{chava} and, more recently, to map certain models of active matter through a nonlinear extension of the Schr\"odinger equation \cite{wittkowski}. No less intriguing is the perspective offered by Bose-Einstein condensates (BECs), whose dynamics can be captured the Gross-Pitaevskii equation (GPE) known to  describe the ground state of a quantum system of identical bosons 
using a single-particle wavefunction approximation \cite{bradley,MOCZ,PhysRevA.88.033610,luca_laser}.
In this respect, of particular relevance are polaritons, i.e. quasi-particles observed in semiconductors and operating in the strong-coupling regime between bound electron-hole pairs and photons \cite{dang,carusotto}. 
These bosonic particles can spontaneously condense in a phase whose microscopic dynamics has 
been shown to map onto the Kardar-Parisi-Zhang (KPZ) equation \cite{gladilin,deligiannis}.

In this contribution we develop a related yet different analogy, namely a  ``Navier-Stokes-Schr\"odinger'' equation, meaning by this an inverse-Madelung formulation of the Navier-Stokes equation leading to a dissipative Schr\"odinger equation strictly equivalent to a Navier-Stokes fluid (see Fig.\ref{fig1}). This quantum wave representation of a dissipative fluid is built via a shift of the non-linear potential, which includes the quantum Bohm term, removing the quantum pressure, and the laplacian of the phase field, leading to the viscous contribution proportional to the laplacian of the fluid velocity.
The interest towards this description is twofold: On the one hand, the quantum wave formulation may hold interest for describing the dynamics of dissipative quantum fluids; on the other hand the classical analogy could be relevant for quantum computers, for this may permit to simulate fluids using a quantum-mechanical formalism. \cite{MARGIE,EPLQC}. Finally, drawing inspiration from the studies on GPE and polaritons, we also show that a dissipative momentum equation, akin to the Navier-Stokes one, can be obtained by considering an imaginary diffusivity alongside a suitable imaginary potential. We note that our formulations differs from previous ones, such as the Schr\"odinger-Langevin equation \cite{kostin}, where the dissipation stems from  a suitable operator  proportional to the logarithm of the wavefunction and enters via the typical drag term of the Langevin equation, with no scale selectivity in space, a crucial feature of dissipative fluids.

The paper is organized as follows. We initially summarize the calculations  showing how the Madelung equations are obtained from the Schr\"odinger one. In the next section we illustrate how this formalism can be extended to include the correct dissipation contribution via a shifted non-linear potential, while a separate section is dedicated to discussing an alternative solution built upon a complex quantum diffusivity, whose formalism is of relevance for polariton condensates.  
After shortly describing the effects of vorticity, we conclude the paper with some 
remarks on the potential perspectives of our results for the quantum simulation of classical fluids. 

\section{The Madelung fluid: A recap}

Let us begin by writing the Schr\"odinger equation:
\begin{equation}
\label{SCE}
i \hbar \partial_t \psi = -\frac{\hbar^2}{2m} 
\nabla^2 \psi + V \, \psi , 
\end{equation}
where $\psi(\vec{x},t)$ is the wavefunction at 
position $\vec{x}$ and time $t$, $i$ is the imaginary unit, 
$\hbar$ is the reduced Planck constant, $m$ is the mass, 
$\partial_t$ is the time derivative operator, and 
$\nabla^2$ is the Laplacian operator. Here the potential $V$ is 
assumed to be the sum of two contributions: 
\begin{equation}
V = U + W(|\psi|^2) , 
\end{equation}
where $U(\vec{x},t)$ is an external potential and $W(|\psi(\vec{x},t)|^2)$ 
is a nonlinear self-interaction, such as, for instance, in the Gross-Pitaevskii equation. 
Upon dividing Eq. (\ref{SCE}) by $\hbar$ one gets:
\begin{equation}\label{sch_eq}
i \partial_t \psi = - \frac{D}{2} 
\nabla^2 \psi + \Omega \, \psi , 
\end{equation}
where 
\begin{equation}
D = \frac{\hbar}{m}
\end{equation}
is the quantum diffusivity and 
\begin{equation}
\Omega= \frac{V}{\hbar}
\end{equation}
has dimensions of a frequency.
Next we represent the complex wavefunction in eikonal form 
\begin{equation}
\psi = R \, e^{i s} , 
\label{euno}
\end{equation}
where $R(\vec{x},t)$ is the real amplitude and 
$s(\vec{x},t)$ is the phase field, which can be also written as $s(\vec{x},t)=S(\vec{x},t)/\hbar$ with $S(\vec{x},t)$ an action field. 
Madelung \cite{Madelung1,Madelung2} suggested to interpret the Schr\"odinger field as the complex field of fluid with local number density 
\begin{equation}
\rho = R^2 
\label{edue}
\end{equation}
and local velocity 
\begin{equation}
\vec{u} = D \vec{\nabla} s. 
\label{etre}
\end{equation}
By inserting Eqs. (\ref{euno}), (\ref{edue}), and (\ref{etre}) into Eq. (\ref{SCE}) it is straightforward to obtain two coupled 
equations associated to the real and imaginary part of Eq. (\ref{SCE}). These equations are the continuity equation 
\begin{equation}
\partial_t \rho + \vec{\nabla} \cdot 
( \rho \vec{u} ) = 0 
\label{continuity}
\end{equation}
and the Euler-like equation 
\begin{equation}
\partial_t \vec{u} + \vec{\nabla} (\frac{u^2}{2}+\frac{q^2}{2} + D\Omega) = \vec{0} ,
\end{equation}
where we have defined
\begin{equation}
\frac{q^2}{2} \equiv \frac{Q}{m} = -\frac{\hbar^2}{2m^2} \frac{\nabla^2 R}{R}
=-\frac{D^2}{2}\frac{\nabla^2 R}{R} 
= -\frac{D^2}{2}\frac{\nabla^2 \sqrt{\rho}}{\sqrt{\rho}} .
\end{equation}
Note that since the quantum potential $Q$ is signed, namely positive in regions of 
negative $R$ curvature and vice-versa, $q^2$ is also signed. We use this notation
to emphasize that $q$, be it real or imaginary, has the dimension of a velocity, like $\vec{u}$. 
 
The above relation can be further elaborated by recalling the vector identity
$$
\frac{1}{2}\vec{\nabla} (u^2) = (\vec{u} \cdot \vec{\nabla}) \vec{u} - \vec{u} \times \vec{\omega},
$$
where $\vec{\omega} = \vec{\nabla} \times \vec{u}$ is the fluid vorticity.
Since the flow derives from a gradient, we have  $\vec{\omega} = 0$, so that 
we are left with
\begin{equation}
(\partial_t + \vec{u} \cdot \vec{\nabla}) \vec{u} = \frac{\vec{F}}{m} 
-\vec{\nabla} \left(\frac{Q}{m} + \frac{W}{m}\right) , 
\label{euler}
\end{equation}
where $\vec{F}(\vec{x})=-\nabla U(\vec{x})$ is the force acting on the system 
and $(\partial_t + \vec{u} \cdot \vec{\nabla})$ is the material derivative. 
These are the equations of a compressible, inviscid, irrotational flow subject to the nonlinear self-interaction potential $W(\rho)$ and the nonlinear 
self-interaction quantum potential $Q(\rho)$. 

\section{Madelung fluid with a shifted nonlinear potential} 

\begin{figure}[htbp]
\includegraphics[width=1.0\linewidth]{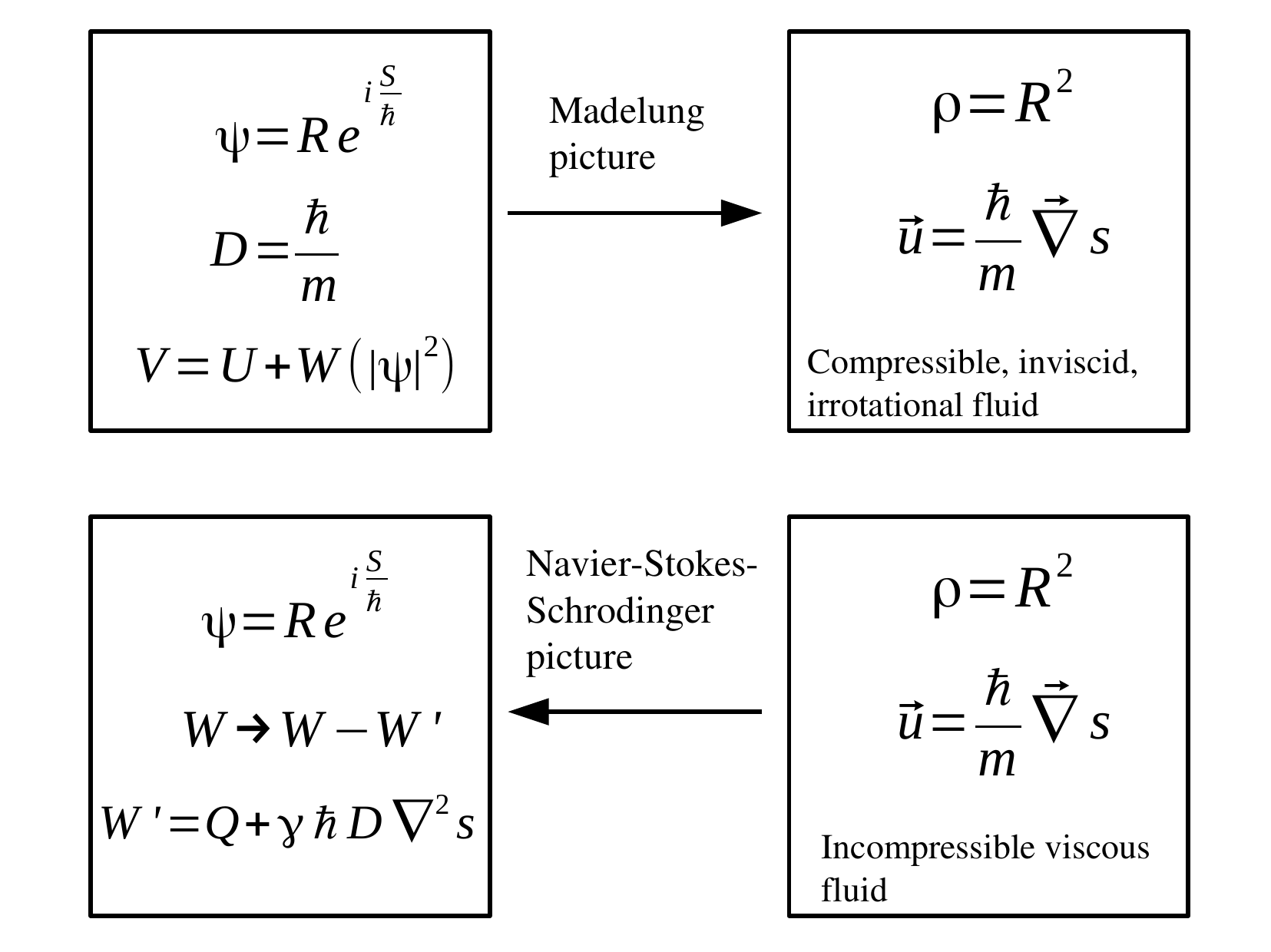}
\caption{Top row: The two quadrants provide a sketch of the Madelung mapping linking the Schr\"odinger equation to  continuity plus momentum equations for a compressible, inviscid, irrotational fluid. Bottom row: The quadrants show the Navier-Stokes-Schr\"odinger mapping proposed in the present paper.}
\label{fig1}
\end{figure}

Here, we consider the case of a dissipative Schr\"odinger equation in which the dissipation is introduced via a shift of the nonlinear potential $W$. The shift is 
\begin{equation}
W \to W - W',
\label{shift}
\end{equation}
where  $W'=Q+\gamma\hbar D \nabla^2 s$, where $\gamma$ is a dimensionless dissipative coefficient which could depend on the density $\rho$. Here, $Q$ removes the quantum potential from Eq. (\ref{euler}) 
while $\gamma\hbar D \nabla^2 s$ introduces a viscous term in 
Eq. (\ref{euler}) (see also Fig.\ref{fig1}). Thus, Eq.(\ref{euler}) becomes 
\begin{equation}
(\partial_t + \vec{u} \cdot \vec{\nabla}) \vec{u} = \frac{\vec{F}}{m} 
-\frac{\vec{\nabla} P}{\rho}  - \frac{\mu}{\rho} \nabla^2 \vec{u} , 
\label{navier}
\end{equation}
taking into account that $\vec{\nabla} (\nabla^2 s)=\nabla^2 ({\vec \nabla} s)$, introducing 
the pressure $P(\rho)$, such that 
$\vec{\nabla} P = (\rho/m)\vec{\nabla} W$, and the shear viscosity 
\begin{equation}
\mu=\gamma D\rho. 
\end{equation}
Quite remarkably, Eqs. (\ref{continuity}) and (\ref{navier}) are nothing but
the Navier-Stokes equations of an incompressible and irrotational fluid. 

Applying the shift of Eq. (\ref{shift}) into Eq. (\ref{SCE}) we get instead 
\begin{eqnarray}
i \hbar \partial_t \psi =  \left[-\frac{\hbar^2}{2m} 
\nabla^2 + U + W(|\psi|^2) + \kappa \frac{\hbar^2}{2m}\frac{\nabla^2 
|\psi|}{|\psi|}\right.\nonumber\\\left. +i\gamma(|\psi|^2)\frac{\hbar^2}{m} \nabla^2 \ln\left(\frac{\psi}{|\psi|}\right) \right] \, \psi , 
\label{mains}
\end{eqnarray}
where the last term on the right hand side stems from the viscous contribution. Note that we have introduced a free parameter $\kappa$ which controls the  transition from quantum ($\kappa=0$) to classical ($\kappa=1$) regimes. Thus, the standard Madelung picture is recovered in the limit $\kappa=\gamma=0$. 

This is the main equation of our paper. We call it Navier-Stokes-Schr\"odinger equation. Indeed, it is not difficult to 
prove that by inserting Eqs. (\ref{euno}), (\ref{edue}), and (\ref{etre}) 
into Eq. (\ref{mains}) one obtains the Navier-Stokes Eqs. (\ref{continuity}) and (\ref{navier}).

It is worth emphasizing that the Navier-Stokes-Schr\"odinger equation introduced in this work is quasi completely ``dequantized'', i.e. devoid of quantum physics effects, as it describes a classical dissipative fluid in quantum mechanical vests. The only quantum memoir is the quantization of the circulation, namely 
\begin{equation}
\oint_{\cal C} \vec{u} \cdot d\vec{x} = 
\frac{\hbar}{m} \oint_{\cal C} \vec{\nabla}s \cdot d\vec{x} = 
\frac{\hbar}{m} 2\pi \, n , 
\end{equation}
due to the fact that the wave function $\psi(\vec{x},t)$ is single valued 
and the phase angle $s(\vec{x},t)$ must be an integer multiple of $2\pi$ along any closed contour ${\cal C}$. If the integer number $n$ is different from zero the fluid displays quantized vortices. Eq. (\ref{mains}) could nonetheless be useful in two respects: first, solve the fluid equations in quantum form on classical computers may prove computationally advantageous as compared to existing numerical methods \cite{minguzzi}. Second, the Navier-Stokes-Schr\"odinger equation may form the basis 
for a new class of quantum algorithms to simulate classical fluids on quantum computers.

\section{An alternative formulation}

Drawing from the recent literature on the Gross-Pitaevskii equation \cite{gladilin,deligiannis}, a dissipative Schr\"odinger equation can be obtained by making both the quantum diffusivity and the potential complex, namely $D_{\gamma} = D(1+i\gamma)$ and $\Omega=\Omega_1+i\Omega_2$, where $\Omega_1$ and $\Omega_2$ are real-valued functions.
The former contribution (extensively studied to develop an analogy between polariton systems and  the KPZ equation \cite{kardar}) generates a dissipative term in the Navier-Stokes equation while the latter one is necessary to restore unitarity, which would otherwise be inevitably lost in the presence of a complex diffusion.

Indeed, taking the imaginary part of Eq.(\ref{sch_eq}) with the extra terms due to imaginary diffusion and potential, one gets
\begin{equation}
\partial_t\rho+\vec{\nabla}\cdot(\rho \vec{u})=\frac{\gamma\rho}{D}(u^2+q^2)+2\rho\Omega_2,
\end{equation}
which shows that the sole imaginary diffusion would contribute a source term leading to loss of unitarity. This issue is overcome by setting
$\Omega_2=-\frac{\gamma}{2D}(u^2+q^2)$, so that the continuity equation is recovered. Note that $\Omega_2$ is not a potential in any conventional sense, as it depends on the fluid velocity, as well as on the gradients of the density. Hence, it is rather to be interpreted
as an adaptive {\it pseudo-potential}, explicitly tailored to absorb non-unitary effects.

Next we consider the real part, which gives

\begin{equation}
(\partial_t+\vec{u}\cdot\vec{\nabla})\vec{u}= -\vec{\nabla}(\frac{Q}{m}+\frac{V}{m})- \frac{\gamma D}{2}\nabla^2 \vec{u}-\gamma\vec{\nabla}(\vec{v}\cdot\vec{u}), 
\end{equation}
 where we have set $\vec{v}=D\vec{\nabla}\rho/\rho$ and $\Omega_1=V/\hbar$.

The above expression recovers the Navier-Stokes equations, provided the 
following conditions are met: 

\vspace{3mm}
{\it i)} $\frac{\vec{\nabla} P}{\rho} = \frac{\vec{\nabla} V}{m}$;

\vspace{3mm}
{\it ii)} $Q=0$; 

\vspace{3mm}
{\it iii)} $\vec{\nabla} (\vec{v} \cdot \vec{u})=\vec{0}$
 
In particular, conditions {\it ii)} and {\it iii)} are both satisfied by letting $\rho = const$,  corresponding to a uniform, incompressible flow.
A softer version of the above could be conceived by requiring that $|q^2| \ll u^2$ and $\vec{v} \cdot \vec{u}=0$. The former is tantamount to stating that the ``kinetic energy'' of the quantum fluctuations  is much smaller than the fluid kinetic energy, a statement of weak inhomogeneity akin to the quasi-incompressible limit of classical fluids. The latter is a statement of orthogonality between $\vec{v}$ and $\vec{u}$. Recalling that $\vec{v} = D \vec{\nabla} \rho /\rho$, the above condition is fulfilled
whenever the density gradient is orthogonal to the fluid velocity, a condition typical of 
two-dimensional incompressible turbulence \cite{TURBO2D}.

\section{Vorticity and gradient flows}\label{secA1}
Given that Navier-Stokes fluids are generally not vorticity-free,  a brief 
comment on vorticity is in order.
Vorticity can actually be supported by gradient flows in the form of singular 
vortices acting as point-like defects causing discrete jumps in the phase field. 
This can be appreciated by computing the circulation $C$ of the velocity on a closed contour,  
$C= \oint_{{\cal C}} \vec{u}\cdot d\vec{x}=2\pi u r_{{\cal C}}=const$, with $u=\omega r_{{\cal C}}$ and $r_{{\cal C}}$  radius of the contour. 
If $C$ is conserved, in the limit $r_{{\cal C}}\rightarrow 0$ one has $u\sim 1/r_{{\cal C}}$ and $\omega\sim 1/r_{{\cal C}}^2$, both singular. 
Since $\vec{u} = D \vec{\nabla} s$, the circulation is $C=r_{{\cal C}} \frac{h}{m}|\nabla s|$. 
This is basically a classical fluid in which the only quantum feature is that the angular momentum per unit mass is quantized in units of $h/m$. 

However, a non-vanishing classical vorticity necessarily requires an extension of the Madelung formulation where the fluid velocity is the gradient of the phase (see Eq.(\ref{etre})). In Ref.\cite{dietrich} it is demonstrated that, 
by introducing a supplementary vector field stemming from the Helmholtz decomposition of the fluid velocity and effectively acting as a magnetic field in a plasma, a mapping between a Scr\"odinger equation of a charged particle moving in this field and a Navier-Stokes equation of a dissipative rotational fluid can be actually built.  It is finally worth mentioning that the vorticity can be also introduced in the realm of quantum  mechanics by invoking a quaternion form of the two-component Schr\"odinger-Pauli equation, which also includes a source term that can  eventually mimic dissipation via a spin-dependent forcing contribution \cite{tao,meng}. However, this term is signed and does not take the form of the Navier-Stokes dissipation.

\section{Conclusions}

In this paper we have shown that shifting the non-linear potential of a term proportional to the sum of the quantum potential plus the laplacian of the phase leads to a generalized Schr\"odinger equation which maps a Navier-Stokes equation of an incompressible dissipative fluid.  
Furthermore, higher order dissipative terms could be readily included by shifting the non-linear potential with even powers of the laplacian. 
Although formally irrotational, this Navier-Stokes equation can support vortices 
emerging from phase singularities, while a classical vorticity would necessarily require 
an additional vector field modifying the structure of the Madelung fluid velocity.
Making the quantum diffusivity complex represents an alternative route to account 
for the dissipation. 
Although this is known to basically destroy  the quantumness of the system \cite{BERRY}, 
the unitarity can be {\it formally} circumvented by 
introducing  an {\it ad hoc}  imaginary pseudo-potential. 
This is not going to restore quantumness in any physical sense, but serves the purpose of 
casting a classical problem, Navier-Stokes fluid dynamics, in quantum mechanical form.  
Unlike the previous one, this formulation is however restricted to the case where 
density is i) constant in space and time and ii) orthogonal to the fluid velocity.

It is important to stress that, at finite temperature and below the critical temperature of the superfluid-to-normal phase transition, quantum fluids are characterized by both quantized vortices and viscosity. In the two-fluid model of Landau \cite{Landau}, the viscousless
and irrotational superfluid component is responsible of quantized vortices while the normal component takes into account the viscosity. Here we are proposing a quite general single-fluid model
which can be also used for quantum fluids. Our formulation has some similarities with the
dissipative nonlinear Schr\"odinger equation adopted by some authors
(see, for instance, \cite{Choi,Tsubota,Nikolaieva}) to study numerically the formation and dynamics of quantized vortices in superfluid liquid 4He or in Bose-condensed atomic quantum gases.
This peculiar nonlinear Schr\"odinger equation contains an imaginary dissipative term but also a chemical potential that fixes the total number of particles when the fluid eventually approaches a stationary configuration.

Besides an interest as a formal connection between quantum physics and dissipative fluid dynamics, both approaches discussed in the present paper open intriguing perspectives for 
simulating incompressible Navier-Stokes equations on modern quantum computers. 
While the first method proposes a surprising simple solution based on a shifted non-linear potential, the second one builds the analogy at the price of loss of unitarity in the absence of the imaginary pseudo-potential. Lack of unitarity would affect the GPE-KPZ analogy as well, but since this 
analogy has received experimental confirmation  over the last few years, we are led 
to speculate that there must be a region of experimental 
parameters such that the violation of unitarity can be neglected. 
This may open the intriguing perspective of using polaritons for the quantum 
simulation of the incompressible fluids \cite{QPOL}.

\section*{Acknowledgments}
SS acknowledges illuminating discussions with Prof. Michael Berry. He also acknowledges financial support from the Flatiron Institute at the  Simons Foundation, where part of this work was performed. LS is partially supported by the European Quantum Flagship  Project ``PASQuanS 2'', by the European Union-NextGenerationEU within  the National Center for HPC, Big Data and Quantum Computing  [Project No. CN00000013, CN1 Spoke 10: ``Quantum Computing''], 
by the BIRD Project ``Ultracold atoms in curved geometries'' of the  University of Padova, by ``Iniziativa Specifica Quantum'' of INFN, and by PRIN  Project ``Quantum Atomic Mixtures: Droplets, Topological Structures, and Vortices'' of the 
Italian Ministry for Universities and Research.

\end{document}